\documentclass[12pt,preprint]{aastex}

\usepackage{epsfig}
\usepackage{color,colordvi}

\newcommand{\BV}{{\bf BV}}
\newcommand{\B}{{\bf B}}
\newcommand{\BdotR}{$\mathbf{B} \cdot \mathbf{R}$}
\newcommand{\HeI}{He$^\circ$}
\def\nHI{n(H$^\mathrm{o})$}
\newcommand{\Qrm}{$Q/m$}
\newcommand{\Pol}{$P$}
\newcommand{\PA}{$\theta_\mathrm{PA}$}
\newcommand{\dPA}{$\Delta \theta_\mathrm{PA}$}
\newcommand{\PAcel}{{$\theta_\mathrm{cel}$}}
\newcommand{\PAgal}{{$\theta_\mathrm{gal}$}}
\newcommand{\PAismf}{{$\theta_\mathrm{ismf}$}}
\newcommand{\glon}{$\ell$}
\newcommand{\glat}{$b$}
\newcommand{\elon}{$\lambda$}
\newcommand{\elat}{$\beta$}
\newcommand{\deeg}{$^\circ$}
\newcommand{\cc}{cm$^{-3}$}

\newcommand{\microG}{$\mu$G}
\def \kms {${\rm km~s}^{-1}$}
\def\cmtwo{cm$^{-2}$}
\newcommand{\IBEX}{\emph{IBEX}}
\newcommand{\Voyager}{\emph{Voyager}}
\newcommand{\Ulysses}{\emph{Ulysses}}
\newcommand{\V}{{\bf V}}
\newcommand{\R}{{\bf R}}
\newcommand{\Bnofil}{B$_\mathrm{CLIC}$}

\newcommand{\Bfil}{$B_\mathrm{filament}$}

\slugcomment{v1 submitted to ApJ 1/19/15} 

\shorttitle{Evidence for dust filament in outer heliosheath}
\shortauthors{Frisch et al.}

\begin{document}

\title{EVIDENCE FOR AN INTERSTELLAR DUST FILAMENT IN THE OUTER HELIOSHEATH}
\author{P. C. Frisch}
\affil{Department of Astronomy and Astrophysics, University of Chicago,
Chicago, IL  60637, USA}
\author{B-G Andersson}
\affil{SOFIA, USRA, CA, USA}
\author{A. Berdyugin and V. Piirola}
\affil{Finnish Centre for Astronomy with ESO, University of Turku, Finland}
\author{H. O. Funsten}
\affil{Los Alamos National Laboratory, Los Alamos, NM, USA}
\author{A. M. Magalhaes and D. B. Seriacopi}
\affil{ Instituto de Astronomia, Geofisica e Ciencias Atmosfericas, Universidade de Sao Paulo,
  Brazil}
\author{D. J. McComas\altaffilmark{1}}
\affil{Southwest Research Institute, San Antonio, TX, USA}
\altaffiltext{1}{Also at University of Texas, San Antonio, TX, USA}
\author{N. A. Schwadron}
\affil{Space Science Center, University of New Hampshire, Durham, NH, USA}
\author{J. D. Slavin}
\affil{Harvard-Smithsonian Center for Astrophysics, Cambridge, MA, USA}
\and
\author{S. J.  Wiktorowicz}
\affil{Department of Astronomy, University of California at Santa Cruz, Santa Cruz, CA, USA}

\begin{abstract}

A recently discovered filament of polarized starlight that traces a
coherent magnetic field is shown to have several properties that are
consistent with an origin in the outer heliosheath of the heliosphere:
(1) The magnetic field that provides the best fit to the polarization
position angles is directed toward \glon=357.3\deeg, \glat$=17.0^\circ
~(\pm 11.2^\circ )$; this direction is within $6.7^\circ \pm
11.2^\circ $ of the observed upwind direction of the flow of
interstellar neutral helium gas through the heliosphere.  (2) The
magnetic field is ordered; the component of the variation of the
polarization position angles that can be attributed to magnetic
turbulence is $ \pm 9.6^\circ$.  (3) The axis of the elongated
filament can be approximated by a line that defines an angle of
$80^\circ \pm 14^\circ$ with the plane that is formed by the
interstellar magnetic field vector and the vector of the inflowing
neutral gas (the \BV\ plane).  We propose that this polarization
feature arises from aligned interstellar dust grains in the outer
heliosheath where the interstellar plasma and magnetic field are
deflected around the heliosphere.  This interpretation suggests that
the polarization is seen where stream lines of the flow and the draped
interstellar magnetic field lines are approximately parallel to each
other and perpendicular to the sightline.  An ordered magnetic field
is required so that grain alignment is not disrupted during the
interaction.  The filament location is consistent with dust plumes
previously predicted to form around the heliosphere.  The proposed
outer heliosheath location of the polarizing grains can be tested with
three-dimensional models that track torques on asymmetric dust grains
as they propagate through the heliosheath plasma, and using these
models to evaluate grain alignment and the asymmetric extinction of
the grains.
\end{abstract}


\newpage
\pagebreak

\section{INTRODUCTION}  \label{sec:intro}

The Sun's heliosphere and the astrospheres of other stars sweep up
interstellar gas and dust as they travel through interstellar clouds.
Infrared data show that dusty bow waves, bow shocks, and shells are
common around other stars \citep{Peri:2012bowshocksurvey}.
Three-dimensional MHD models of the interactions between interstellar
dust grains and the heliosphere predict that dust plumes composed of
sub-micron grains with high charge-to-mass ratios form around the
heliosphere \citep{SlavinFrisch:2012}.  Detection of such a dusty
shell around the heliosphere using infrared measurements would be
challenging because of the difficulty in separating background
emissions from a heliospheric feature.

The center of the ``ribbon'' arc of energetic neutral atoms (ENAs),
discovered by the Interstellar Boundary Explorer (\IBEX) mission,
traces the interstellar magnetic field draping over the heliosphere
\citep{McComas:2009sci,Schwadron:2009sci,Funsten:2009sci,Funsten:2013,Funsten:2014,Heerikhuisen:2014}.
The extraordinary circularity and symmetry of the ribbon supports a
hypothesis that the interaction between the heliosphere and the
interstellar medium (ISM) involves an ordered magnetic field.
{Charged interstellar dust grains interacting with the ordered field
  may polarize starlight and provide an alternate method for sampling
  a dusty shell around the heliosphere.}

The Sun is traveling through an interstellar medium (ISM) containing
partially ionized interstellar gas with low average densities, a
magnetic field, and dust \citep{Frisch:2011araa}.  The electron
density, $\sim 0.03 - 0.1$ \cc, and magnetic field strength, $\sim 3$
\microG, in the surrounding interstellar gas are typical of the
magnetoionic medium that contains a structured magnetic field and
fills $ 10\% - 40\%$ of space
\citep{Haverkorn:2010magnetoionic,Haffner:2009review}.  About
0.5\%--0.7\% of the local interstellar mass density is in charged dust
grains \citep{SlavinFrisch:2008,Krueger:2014mass}.  Mapping the
structure of the nearby magnetic field requires high-sensitivity
measurements of starlight that is polarized by asymmetric dust grains
aligned with respect to the foreground interstellar magnetic field
(ISMF).

This is the fourth study in a global survey of the polarizations of
light from nearby stars in order to map the ISMF within 40 pc and
compare it with the direction of the ISMF shaping the heliosphere
\cite[][Papers I, II,
  III]{Frisch:2010ismf1,Frisch:2012ismf2,Frisch:2015ismf3}. The survey
includes stars within $\sim 90^\circ$ of the galactic center.  A
magnetic filament was identified in these polarization data through
the linear rotation of their polarization position angles with
distance (Paper III).  The filament forms a band over 100\deeg\ long
and $\sim 15^\circ$ wide in the sky, and extends to within 10 pc of
the Sun.  Paper III discusses a possible association of this filament
with the Loop I superbubble.  Evidence that this magnetic filament may
be associated with the heliosphere is the topic of this paper.

In the diffuse interstellar medium where the ISMF direction is also
traced by the Faraday rotation of radio waves and the polarization of
synchrotron emission, the optically asymmetric interstellar dust
grains that linearly polarize starlight are aligned so that plane of
polarization is parallel to the ISMF direction, and is strongest where
\BdotR=0 for magnetic field \B\ and radial sightline
\R\ \citep[e.g.][]{DavisBerge:1968,Lazarian:2007rev,Andersson:2014araa}.
The polarization position angles\footnote{The polarization position
  angle is the angle between the plane of polarization of the E-vector
  and the north pole, and labeled positive for rotations of the
  postion-angle meridian toward larger longitudes.}  of the survey
data serve as a method for evaluating ISMF direction that is traced by
the polarization the data.  In Paper III, the filament star
polarizations were found to track a separate ISMF direction than the
other polarization data in the sample.  After the filamanet subset of
data was excluded from the analysis of the ISMF direction, the
dominant galactic magnetic field in the immediate solar vicinity was
found to have a magnetic pole located at
$\ell=36^\circ,~b=49^\circ,~(\pm 16^\circ)$ \citep[Paper
  III,][]{Frisch:2014icns}.  This dominant local ISMF direction agrees
to within 1-sigma with the ISMF direction traced by the weighted-mean
center of the highly circular \IBEX\ ribbon arc \citep{Funsten:2013}.
The nominal pole directions differ by $7.6^\circ \pm 16.2^\circ$.

The velocity vectors of interstellar dust and gas flowing into the
inner heliosphere are the same.  Dust arrives from
$\ell=8^\circ,~b=14^\circ~(\pm 13^\circ)$
\citep{Frisch:1999,Belheouane:2012stereodust,Mann:2010araa} at a
velocity of --24.5 \kms\ \citep{KimuraMann:2003vel}.  Neutral helium
arrives from the direction \glon=3\deeg, \glat=16\deeg, at a velocity
of --26
\kms\ \citep{Bzowski:2014ulysses,Witte:2004,Moebius:2012isn,Bzowski:2012isn,Wood:2014ulysses,KatushkinaWood:2014ulysses,McComas:2014warm}.
There is a deficit of low-mass interstellar grains ($<10^{-13}$ gr) in
the inner heliosphere compared to the parent interstellar population
that is attributed to the action of Lorentz forces on grains with high
charge-to-mass ratios in the outer heliosphere and heliosheath regions
\citep{KimuraMann:1998charge,Frisch:1999,LindeGombosi:2000,SlavinFrisch:2012,Sterken:2012}.
Three-dimensional models of the trajectories of interstellar dust
grains interacting with a three-dimensional MHD heliosphere model
predict that these excluded grains create shells or plumes of dust
around the heliosphere \citep{SlavinFrisch:2012}.

\section{PROPERTIES OF THE MAGNETIC FILAMENT} \label{sec:description}

\begin{figure}[t!]
\begin{centering}
\includegraphics[height=.4\textheight]{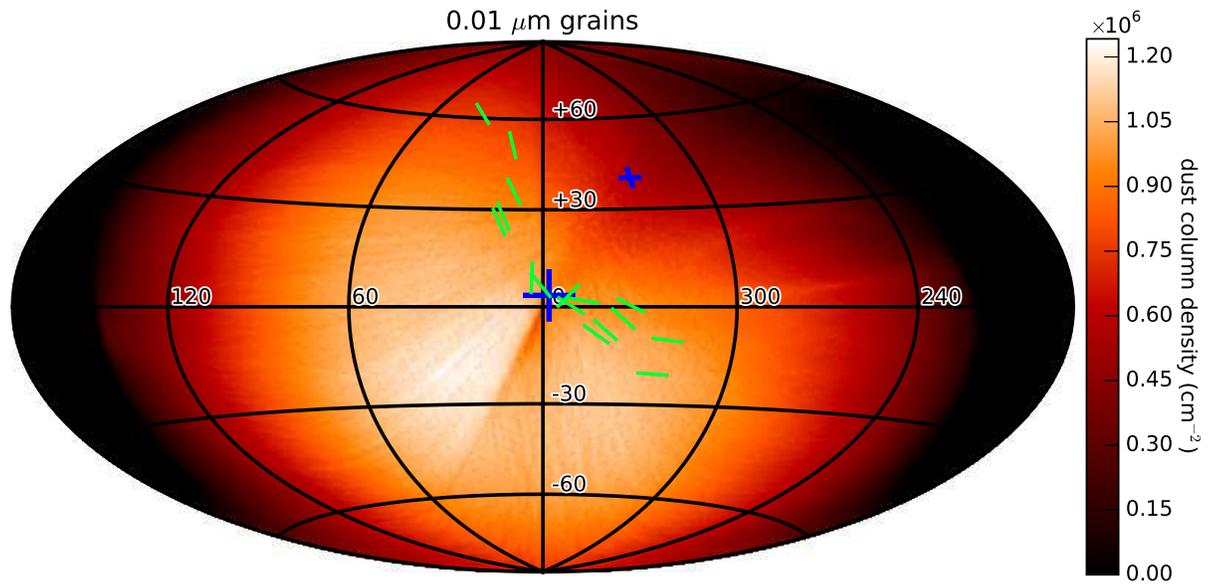}
\end{centering}
\caption{The filament star polarizations (green bars) are shown
  plotted against the modeled column densities of radius 0.01
  \micron\ interstellar dust grains interacting with the
  heliosphere. The small and large blue crosses indicate the direction
  of the \IBEX\ ribbon ISMF and inflowing interstellar gas.  The
  horizontal axis coincides with the ecliptic plane and the image is
  centered on the heliosphere nose defined by the inflowing ISM.  The
  color coding shows the column density for 0.01 dust micron grains
  out to a distance of 400 AU.  Color coding is based on the true
  column density (\cmtwo) for the grains for a gas-to-dust ratio of
  150.  The dust distribution is from the dust trajectory models of
  \citet{SlavinFrisch:2012}}
\label{fig:js}
\end{figure}

An ordered component of the local ISMF within 40 pc was originally
noticed in Paper II, using the high-sensitivity polarization data set
acquired with PlanetPol \citep{planetpol:2010}.  The distinguishing
characteristic of this ordered ISMF was that the polarization position
angle in the equatorial coordinate system (\PAcel) rotated linearly
with the distance of the star.  Using an expanded set of stars that
included the PlanetPol data, we found a linear relation between the
polarization position angle in galactic coordinates, \PAgal, and the
star distance (Paper III).  The stars separated into two groups that
had slightly different distance dependences for \PAgal, neither of
which agreed with the gradient found from \PAcel\ in Paper II.  This
difference was a clue that the apparent relation between polarization
position angle and distance was spurious, and that another factor
organized these polarization position angles.  The stars showing the
regular variation of position angle, \PA, with distance were found to
trace an extended filamentary feature of linearly polarized starlight
that extends over 100\deeg\ in the sky. The polarization data tracing
this filament are listed in Table \ref{tab:filament}, and plotted in
Fig. \ref{fig:js} in a nose-center ecliptic coordinate
system.\footnote{The heliosphere nose is defined by the direction of
  inflowing neutral interstellar helium atoms, \elon=255\deeg,
  \elat=5\deeg, or \glon=4\deeg,
  \glat=15\deeg\ \citep{McComas:2014warm}}.

In order to search for the organizing principle of the filament, data
on three new stars in the region of the filament were collected with
DiPol-2 at the UH88 telescope \citep{Piirola:2014spie}.  These new
data are also listed in Table 1. These polarization data that define
the filament were acquired using different instruments at seven
observatories; the mean errors of the measurements are not uniform
between these different data sets.  This extended set of polarization
data of sixteen stars is used here to search for the controlling
factor that creates the alignment of the filament polarizations.
Several stars have multiple measurements; all data were used in this
analysis.  This selection of data is relatively unbiased in the sense
that only the three stars observed most recently by DiPol-2, HD
153631, HD 141937, and HD 145518, were observed for the purpose of
confirming the filament properties.

The first test performed on the filament polarizations is to determine
the direction of the ISMF associated with these polarizations.  The
best-fitting ISMF to the polarization data in Table \ref{tab:filament}
was obtained with the method in Papers II and III, where it is assumed
that the ISMF direction can be described by a great circle in the sky,
and that the best-fitting ISMF direction is the one that minimizes the
ensemble of sines of the polarization position angles referenced to
the ISMF great circle.\footnote{In the context of the software, this
  assumption is implemented by rotating the coordinate system into a
  1-degree grid covering the sky to find the direction that
  best-matches the rotated polarization position angles.  No
  assumption is made about the three-dimensional distribution of the
  ISMF since the polarizing dust screens can be anywhere in the
  sightline.  See \citet{Frisch:2015ismf3} for more details.}  A
polarization position angle that is perfectly aligned with the ISMF
direction will have a position angle of either 0\deeg\ or
180\deeg\ when referenced to the ISMF direction, so that the sine of
the polarization position angle, \PAismf, is zero.  The polarization
strengths are not used as the basis for determining the ISMF
direction.

Testing this set of position angles against all possible ISMF
directions, using a 1-degree grid and the weighted means of the
position angles to evaluate the minimum, yields a best-fitting ISMF
direction to the polarizations of the filament stars that is toward
\glon=357.3\deeg, \glat=17.0\deeg\ ($\pm 11.2^\circ$).  The
uncertainties on this direction are obtained from the standard
deviation of the polarization position angles calculated with respect
to the magnetic field frame, \PAismf, that corresponds to the
best-fitting magnetic field pole.  An estimate of the magnetic
turbulence of this ordered field is obtained by assuming that the
uncertainty of 11.2\deeg\ is composed of the root-mean-square of the
turbulent field component and the mean uncertainty of the position
angle measurements, \dPA=5.6\deeg, so that the ``turbulent'' component
of the position angles is approximately $ \pm 9.6^\circ$.

The surprising result of best-fitting ISMF direction to the filament
polarizations is that it is within $6.7^\circ \pm 11.2^\circ$ of the
upwind velocity vector of the primary interstellar neutral helium flow
into the heliosphere, which arrives from the direction of
\glon=3\deeg, \glat=16\deeg\ \citep[or \elon=255\deeg,
  \elat=5\deeg\ in ecliptic coordinates, e.g.][]{McComas:2014warm}.
The coincidence between the directions of ISMF vector and the velocity
vector of neutral interstellar gas flowing into the heliosphere
suggests that the filament may be related to the heliosphere, and that
the filament axis may trace a feature of the heliosphere.

\begin{figure}[t!]
\begin{center}
\includegraphics[height=.5\textheight]{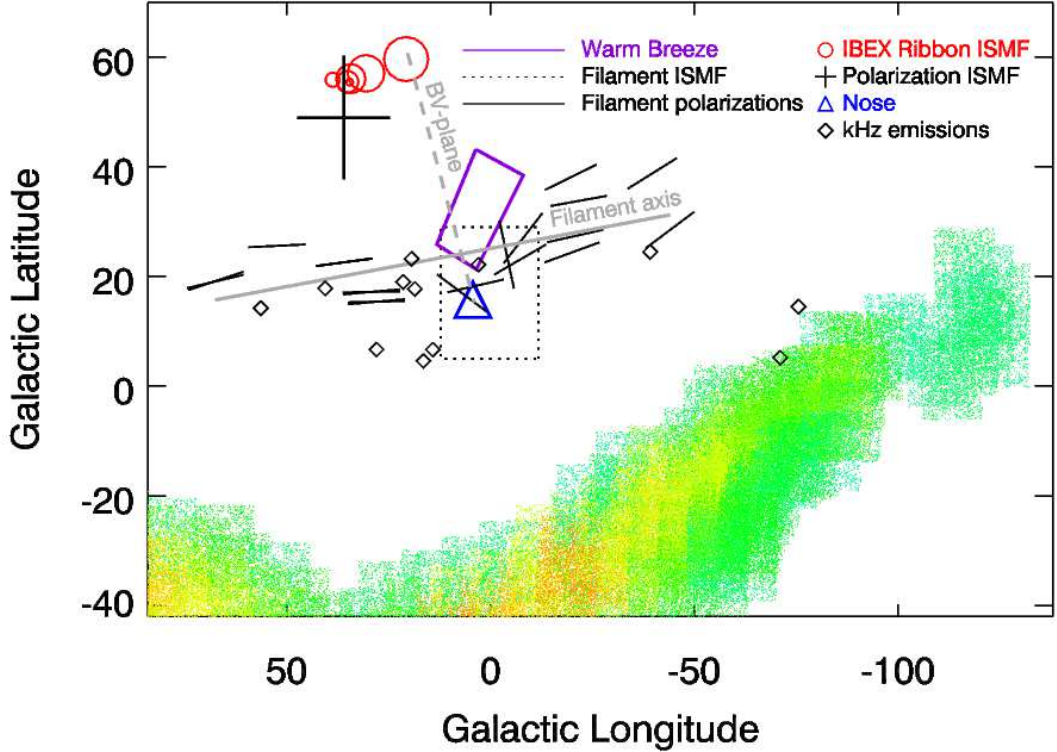}
\end{center}
\caption{The polarizations of the filament stars are compared to
  markers of the outer heliosheath region or beyond.  Those
  polarizations, and the best-fitting magnetic field direction to
  those polarizations \Bfil, are plotted in galactic coordinates as
  black bars and the dotted-line box (which outlines the uncertainty
  range).  The uncertainties on \Bfil\ overlap the upwind direction of
  the primary interstellar \HeI\ atoms (blue triangle) and the
  uncertainty range of the inflowing warm \HeI\ breeze (purple box,
  solid line).  The ISMF directions from the centers of the
  \IBEX\ ribbon arc for the five energy bands of \IBEX-Hi
  \citep{Funsten:2013} are plotted as red open circles, with the size
  of the circle increasing with energy.  The black diamonds represent
  the locations of the low frequency outer heliosheath 2--3 kHz
  emissions reported in \citet{KurthGurnett:2003}.  The magnetic field
  direction that dominates polarization data \citep[\Bnofil, large
    black cross,][]{Frisch:2015ismf3} overlaps the ISMF found from the
  \IBEX\ ribbon.  The dashed and solid gray lines show the locations
  of the \BV\-plane and the axis of the filament, respectively. The
  lines enclose an angle of $80^\circ \pm 14^\circ$ and are
  perpendicular to each other. The fluxes of the \IBEX\ ribbon for the
  $\sim 1.1$ keV energy band are plotted on the color scale, with the
  highest fluxes plotted as red and the lowest fluxes plotted in green
  \citep[see Fig. 4 of ][]{Schwadron:2014sep}.}
\label{fig:heliosheath}
\end{figure}

The major axis of the filament axis can be approximated by fitting a
line through the galactic coordinates of the stars.  The best-fitting
line corresponds to $\ell = 25.118^\circ~(\pm 1.258^\circ) +
0.139^\circ ~(\pm 0.037^\circ)~\times~\delta \ell $, with a reduced
$\chi^2$ of 1.14; here $\delta \ell$ is the difference between the
star galactic longitude and the end point of the polarization filament
at \glon=67\deeg\ (Table 1).  The long axis of the filament is shown
as the gray line in Fig. \ref{fig:heliosheath}.

The angle between the filament axis and the \BV\ plane of the
heliosphere is $80^\circ \pm 14^\circ$, so that the filament axis and
\BV\ plane are perpendicular.  The \BV\ plane, projected as the gray
dashed line in Fig. \ref{fig:heliosheath}, is the plane formed by the
ISMF (\B) and neutral interstellar wind velocity (\V) vectors that
shape the heliosphere
\citep{Pogorelov:2009ribbon,OpherBibi:2009nature}.  For the magnetic
field direction (\B) we use the center of the \IBEX\ ribbon arc at 4.3
keV \citep{Funsten:2013}.  At this energy ENAs will have longer mean
free paths than those at lower energies, so that this direction
extends further into the ISMF than the directions from lower
energies. The interstellar velocity (\V) is given by the velocity
vector of neutral interstellar helium flowing through the heliosphere
\citep{McComas:2014warm}.  Ribbon models indicate that the ribbon
center is within 5\deeg\ of the true ISMF direction for likely field
strengths of 3 \microG\ \citep{Heerikhuisen:2014}.

A range of heliospheric data indicate that the \BV\ plane plays a
fundamental role in organizing the interaction between the ISM and the
heliosphere.  Charge exchange coupling between interstellar neutrals
and ions cause an offset of the neutral gas flow along a direction
that is parallel to the \BV\ plane; examples are the
4.7\deeg\ deflection the interstellar hydrogen flow from the benchmark
\HeI\ flow direction \citep{Lallement:2010soho}, and the offset of the
secondary helium population \citep[a "warm
  breeze",][Fig. \ref{fig:heliosheath}]{KubiakBzowski:2014breeze}.
The \BV\ plane also plays an organizing role in the spectral
dependence of the ribbon symmetry, with ENAs \citep{Funsten:2014}, and
organizes the heliosphere asymmetries (Fig. \ref{fig:HP}).

Figure \ref{fig:js} plots the filament polarizations over a simulation
of the column densities of 0.01--1.0 \micron\ interstellar dust grains
within 400 AU from \citet[][S12]{SlavinFrisch:2012}.  The figure is an
ecliptic projection centered on the nose of the heliosphere, defined
by the inflowing interstellar helium gas.
\citet[][S12]{SlavinFrisch:2012} modeled the trajectories of
interstellar compact silicate grains through the 3D MHD heliosphere
models of \citet{Pogorelov:2009ribbon}.  These models were
successfully used by \citet{Schwadron:2009sci} to show that the
\IBEX\ ribbon appears in sightlines that are perpendicular to the ISMF
draping over the heliosphere, i.e. where \BdotR=0 for magnetic field
vector \B\ and radial direction \R.  The models include an ISMF
direction that is within $\sim 17^\circ$ of the interstellar field
direction traced by the weighted center of the \IBEX\ ribbon
arc.\footnote{The galactic coordinates of the weighted mean center of
  the \IBEX\ ribbon arc are $\ell=36.7^\circ \pm
  2.1^\circ,~b=56.0^\circ \pm 0.6^\circ$
  \citep{Funsten:2013,Frisch:2014icns}.}  Densities of 0.01
\micron\ interstellar grains within $\sim 400$ AU of the Sun are shown
in Figure \ref{fig:js} based on these models.  The filament crosses
the heliosphere nose at the plot center and reaches high latitudes on
the port side of the heliosphere at HD 172167, and low latitudes on
the starboard side of the heliosphere at HD 119756 (borrowing nautical
terminology\footnote{In nautical usage, port and starboard are
  referenced to the bow of a ship.  Using an xyz right-handed
  cartesian coordinate system, the x-axis is directed towards the bow,
  the y-axis is directed toward the port side and the z-axis points
  "up" toward the opposite direction of the gravitational force.  By
  analogy, for the heliosphere the nose-direction of the heliosphere
  corresponds to the x-axis, the y-axis is directed toward the port
  side of the heliosphere, and the z-axis, pointing north, completes
  the right-handed coordinate system.} ) The \BV\ plane controls the
symmetry of the cloud of interstellar dust grains interacting with the
heliosphere, and the asymmetry of the polarization filament about the
\BV\ plane is evident.  The filament is slightly more twisted toward
the vertical direction in the north and port side of the heliosphere
compared to the starboard side in this simulation.

\section{DISCUSSION}

\subsection{Possible Origin of Filament Polarizations in Outer Heliosheath}

The agreement between the ISMF direction of the filament polarizations
and the velocity vector of interstellar material flowing through the
heliosphere, the perpendicularity between the \BV\ plane and the
filament axis, and the low level of magnetic turbulence in the
filament, suggest that the filament polarizations may be associated
with the deflection of charged grains, the ISMF, and interstellar
plasma around the heliosphere in the outer heliosheath regions.

Interstellar nanograins with large charge-to-mass ratios (\Qrm) are
deflected in the outer and inner heliosheath regions, and throughout
the heliosphere
\citep{Frisch:1999,KimuraMann:1999,LindeGombosi:2000,Krueger:2014mass,CzechowskiMann:2003b,MannCzechowski:2004,SlavinFrisch:2012,Sterken:2012}.
Most grain trajectory simulations have assumed compact grains.  Tiny
grains are subject to the small particle effect, where the efficiency
of secondary ejection of electrons is enhanced
\citep{KimuraMann:1998charge}.  Porous or fluffy grains that consist
of aggregates of tiny grains are also subject to this effect, which
leads to higher grain charges and increases the average mass of the
outer heliosheath grains above the predicted levels for compact grains
\citep{Ma:2013fluffyinism}.  These deflected grains are candidates for
the polarizing dust grains in the filament.

The mass distributions of the interstellar dust grains observed within
5 AU of the Sun exhibit a deficit of low mass grains, $<10^{-13}$ gr
\citep{Frisch:1999,Landgrafetal:2000,Krueger:2014mass}, when compared
to the nominal power-law interstellar distribution of
\citet{MRN:1977}.  The deficit of small grains is attributed partly to
the ``filtration'' of grains with large values of \Qrm\ that couple to
the interstellar plasma and magnetic field that is swept around the
heliopause
\citep{Frisch:1999,KimuraMann:1999,LindeGombosi:2000,Krueger:2014mass,CzechowskiMann:2003b,MannCzechowski:2004,SlavinFrisch:2012}.

\begin{figure}[t!]
\plottwo{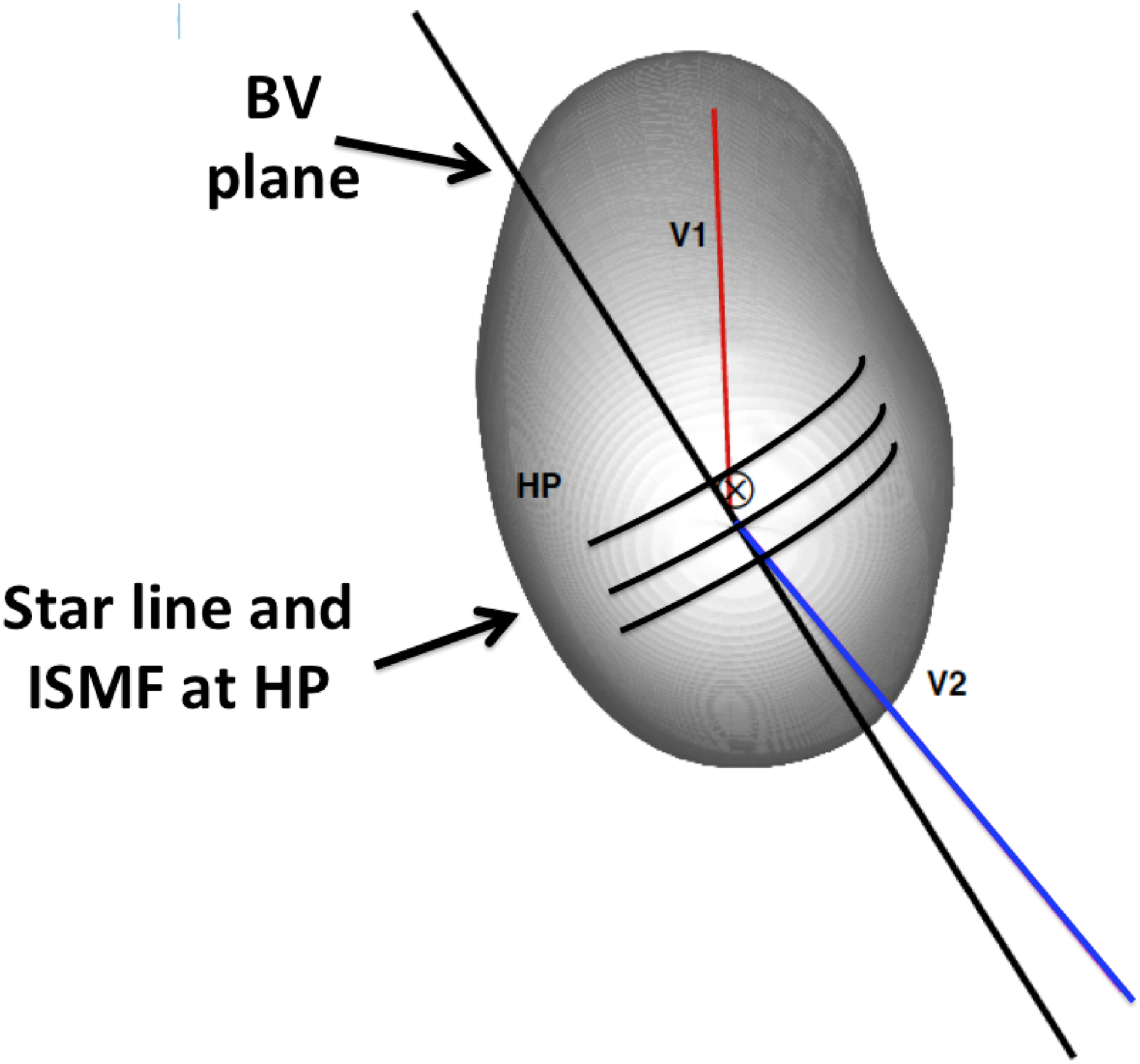}{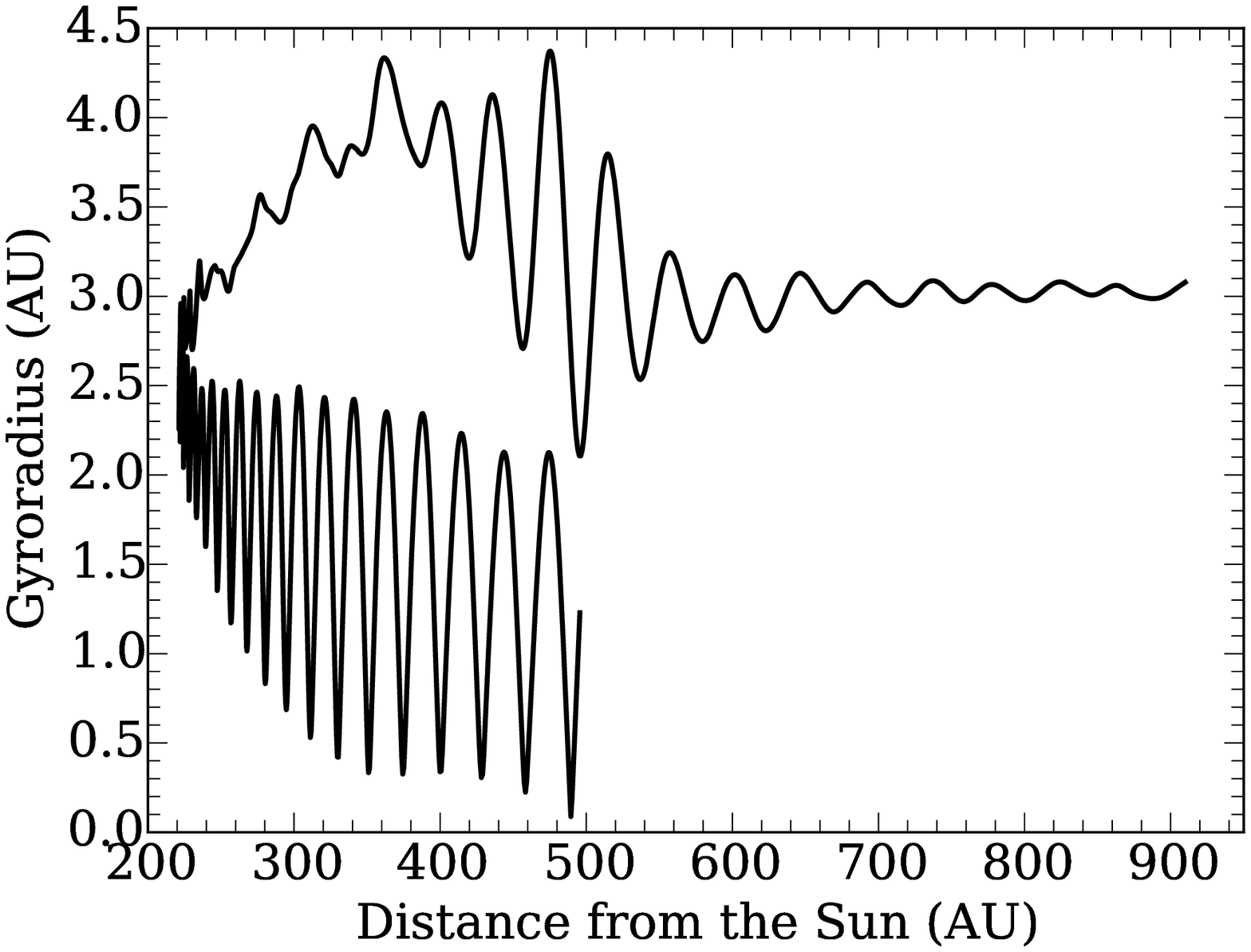}
\caption{Left: A cartoon showing the \BV\ plane and filament star line
  plotted against a frontal view of the heliopause, i.e. as viewed
  from the upwind direction.  The heliopause cartoon is borrowed from
  the MHD model displayed in Fig. 1a of \citep{PogorelovFrisch:2011}.
  The \BV\ plane corresponds to the hydrogen deflection plane
  displayed by \citep{PogorelovFrisch:2011} in Fig. 1a.  The
  orientation of the filament star line is displayed by the draped
  lines.  The filament orientation is perpendicular to the \BV\ plane,
  and the filament magnetic field is directed toward the upwind
  velocity of interstellar neutrals, suggesting that a heliospheric
  origin requires the filament to form where the interstellar dust and
  magnetic field are fully deflected at the heliosphere nose.  Right:
  The modeled trajectory of an interstellar grain such as may generate
  the polarization is displayed on the right, for a compact dust grain
  with radius 0.01 \micron\ approaching the heliosphere nose from the
  upwind direction.  The grain gyroradius varies as the grain charge
  is modified by interactions with the heliospheric radiation and
  plasma components, until the grain is finally deflected around the
  heliopause along with the interstellar plasma and magnetic field
  \citep[see][for modeling details]{SlavinFrisch:2012}.}
\label{fig:HP}
\end{figure}

The perpendicular relation between the filament axis and the
\BV\ plane would be an outcome of the Lorentz force on the grains.
Nanograins experience additional charging in the outer heliosheath due
to collisions with the heated compressed interstellar plasma and
enhanced radiation field, with the maximum differential increase in
the surface potential of the grains occurring for the smallest grains
(Fig. 2 of S12).  S12 showed that the surface potentials of compact
interstellar silicate grains in heliosheath regions are enhanced by an
order of magnitude compared to values in the surrounding ISM.  As the
dust grains approach the heliosphere from the distant ISM, Lorentz
forces act to displace grains into directions that are perpendicular
to the \BV\ plane.  The action of the Lorentz force on grains with
large \Qrm\ the grains naturally explains the orientation of the
filament axis with respect to the \BV\ plane.  Fig. \ref{fig:HP} shows
the gyroradius of a 0.01 \micron\ radius interstellar grain
propagating through the outer heliosheath region, and as the grain is
deflected around the heliopause by the Lorentz force; the grain
trajectory model is based on the simulations in S12.

The $\sim 15^\circ$ width of the filament is not yet explained.  Since
all of the smallest grains are deflected, the filament width must
depend on the detailed configuration of the heliopause, the grain
charging that occurs in the outer heliosheath, and the strength of the
Lorentz force.  The dust plumes in the models of S12 (Fig. 13 of S12,
left) have a limited width that is similar to the filament width, and
a similar spatial orientation perpendicular to the \BV\ plane.  Those
dust plumes therefore provide a plausible model for the polarization
filament location, provided that the efficiency of polarization can be
explained.

If the filament has a heliospheric origin, then the second
characteristic of the filament, that the ISMF is directed toward the
inflow vector of interstellar neutrals, implies that the filament
forms where the interstellar ion and dust streamlines are parallel to
each other and to the localized ISMF direction.  The surrounding
interstellar gas consists of about 23\% ions and 77\% neutral atoms
\citep{SlavinFrisch:2008}.  The collisional mean-free-path in the ISM
approaching the heliosheath regions is $\sim 330$ AU
\citep{SpanglerSavage:2011aip}, and is similar to the thickness of the
plasma density ramp outside of the heliopause
\citep{Zank:2013}.  \footnote{The inner heliosheath thickness found
  from Voyager 1 and \IBEX\ data is $\sim 27$ AU
  \citep{Gurnett:2013sciismvoy,Hsieh:2010enahp}.}

Interstellar ions and neutrals follow different trajectories in the
outer heliosheath.  Ions are deflected around the heliopause, and the
neutrals propagate through the plasma, with some loss due to the
formation of ENAs.  The frozen-in interstellar field becomes fully
deflected where the ions decouple from the inflowing interstellar
neutrals at the heliopause, so that the deflected plasma streamlines
and magnetic field should be directed toward the undeflected neutral
velocity vector.  A correspondence between the ISMF direction and the
\HeI\ flow velocity suggests that the direction of the draped ISMF has
twisted into the direction of the deflected gas flow
\citep[e.g. see][]{OpherDrake:2013}.

An important requirement is that the dust grains remain aligned with
respect to the ISMF direction during the deflection of the dust flow
at the heliopause.  The organized set of filament polarizations must
be tracing an organized magnetic field that has a minimal turbulence.
The recent in situ measurements of the magnetic field in the inner and
outer heliosheath regions with instruments on board \Voyager\ 1
support an ordered field in the outer heliosheath
\citep{Burlaga:2013ismf,Burlaga:2014ismf,Burlaga:2014ismfStDev}.
\Voyager\ 1 crossed into the interstellar magnetic field in mid-2012
and found a more ordered interstellar magnetic field in the outer
heliosheath than had been seen in the inner heliosheath, with less
turbulence, and a maximum standard deviation of the magnetic field
strength that was 3\% of the total field strength.  In contrast, in
the inner heliosheath the maximum standard deviation of the
compressive turbulence (corresponding to variations in magnetic field
strength) was 36\% of the total field strength suggesting that the
inner heliosheath is unlikely to be the origin of the polarizations.

The low level of interstellar field turbulence in the outer
heliosheath found by \Voyager\ 1 suggests that high magnetic pressures
will concentrate the nanograins in the outer heliosheath and increase
the efficiency of the grains to linearly polarize light.  However,
Rayleigh-Taylor instabilities can disrupt the heliopause and allow the
charged interstellar particles to penetrate into the inner heliosheath
\citep{BorovikovPogorelov:2014instability,AvinashZank:2014instability}.
The blunt heliosphere inferred from \Voyager\ data of the neutral
current sheet \citep{Burlaga:2014ismf} and plasma flow
\citep{RichardsonDecker:2014}, and the extended high-pressure region
at the heliosphere nose seen in \IBEX\ data
\citep{McComas:2014fiveyear} and predicted by heliosphere models
\citep{PogorelovFrisch:2011}, is going to affect the draping of the
ISMF over the heliosphere in ways that are not yet fully understood.

A 'cartoon' illustrating the geometric relation between the filament
polarizations and the heliosheath is shown in Fig. \ref{fig:HP}; the
surface of the heliopause is from the MHD model of
\citet[][Fig. 1]{PogorelovFrisch:2011}, where the \BV\ plane and
\Voyager\ trajectories are also shown.  The filament orientation
perpendicular to the \BV\ plane is displayed, with the filament
located close to the heliopause on the upwind side.  The hypothesized
local distortion of the magnetic field direction parallel to the
filament and heliopause is also illustrated.

\subsection{Alignment of the Filament with respect to the \IBEX\ Ribbon }

For starlight that is linearly polarized in the ISM, a polarization
position angle that is aligned with the ISMF direction will yield the
strongest polarizations where the ISMF is perpendicular to the radial
sightline, e.g. where \BdotR=0.  In contrast, the filament
polarizations trace a magnetic field direction that is directed toward
the upwind velocity of neutral interstellar gas instead of the ISMF
direction corresponding to the center of the \IBEX\ ribbon arc, $\sim
48^\circ$ away from the gas velocity.  Our hypothesis is that the
polarization filament is visible where the dust grains are entrained
in the ISMF that has been deflected around the heliosphere, so that
polarizations would be strongest where the field lines are
quasi-perpendicular to the sightline for otherwise-equal conditions.

Comparison between MHD heliosphere models and the \IBEX\ ribbon location
show that the $\sim 1$ keV ribbon is observed in sightlines where the
interstellar magnetic field becomes perpendicular to the sightline as
the field drapes over the heliosphere, i.e. where \BdotR=0
\citep{Schwadron:2009sci,McComas:2009sci,Heerikhuisen:2010ribbon,Chalov:2010ribbon,SchwadronMcComas:2013rr}.
Fig. \ref{fig:heliosheath} shows that the filament polarizations are
roughly parallel to the \IBEX\ ribbon.

There is no consensus on the formation scenario for the \IBEX\ ribbon,
although the original model where the ribbon is viewed along
sightlines perpendicular to the ISMF draping over the heliosphere is
generally accepted \citep[for a review
  see][]{McComasLewisSchwadron:2014}.  Formation from secondary ENA's
upstream of the heliopause is currently the leading contender for
explaining the ribbon.  In these models the ribbon ENAs form upstream
of the heliopause where charge-exchange between interstellar neutral
hydrogen atoms and the pickup-ion plasma, which came from the
neutralized outward moving solar wind, create energetic neutral atoms
(ENAs).  These ENAs propagate back toward \IBEX\ if the ion pitch
angles retain the dominant radial momentum of the original outward
flowing particles.  Models relying on secondary pickup ions to
generate the ribbon place the origin $50-100$ AU upstream of the
heliopause \citep{HeerikhuisenPogorelov:2011} up to several hundred
AU.  Presumably the interstellar ions and the dust grains with small
\Qrm\ will couple more tightly to the deflected ISMF than do the
pickup ions that create the neutral ENAs, so that the polarization
filament would appear at the region where the small \Qrm\ grains are
fully deflected rather than precisely where \BdotR=0.

\subsection{Are Outer Heliosphere Grains enough to Polarize Starlight?} 

Interstellar column densities in the outer heliosheath are expected to
be lower than total column densities toward the nearest stars.  The
gas column densities in the outer heliosheath are only 0.06\%--0.9\%
of the total interstellar column densities toward the nearby upwind
star 36 Oph (6 pc), if the hydrogen-wall and interstellar data in
\citet{Wood36Oph:2000} are used to benchmark this relation.  For an
average hydrogen-wall density of \nHI=0.25 \cc, the three-component
hydrogen-wall model of \citet{Wood36Oph:2000} suggests an outer
heliosheath thickness of $\sim 100$ AU toward 36 Oph. If the filament
polarizations are formed in an outer heliosheath region that is on the
order of 100 AU thick, then either grain alignment is significantly
more efficient in the heliosheath than in the local ISM, or an unknown
factor such as foreground depolarization is reducing the interstellar
polarizations of nearby stars.

The outer heliosheath physical conditions differ from those in the ISM
upstream of the heliosphere.  Although detailed models are required to
evaluate the alignment of grains propagating through the outer
heliosheath, the factors that increase polarization efficiency in this
relatively thin region can be described qualitatively.  These factors
include higher grain surface potentials, enhanced strength and
asymmetry of the radiation field, enhanced magnetic field strengths,
low levels of magnetic turbulence, and the absence of foreground
depolarizations, all when compared to the values in the local ISM.
The presence of fluffy or porous grains, which are more sensitive to
secondary electron ejection than compact grains, further enhancing
grain alignment.

Self-consistent simulations of compact interstellar grains interacting
with the heliosheath regions show that surface potentials of the
grains increase by up to an order of magnitude in the outer
heliosheath, depending on grain radius (S12).  The higher \Qrm\ values
will increase grain charge and the Lorentz force on grains, and
minimize possible collisional disruption of the grain alignment.

Radiative torques play a role in grain alignment in interstellar
clouds \citep{Andersson:2014araa}.
\citet{HoangLazarian:2014radiativetorques} have modeled the
polarizations expected for the very local ISM, taking into account
that radiative torques on optically asymmetric grains play a dominant
role in grain alignment in the warm local ISM where gas column
densities are typically $10^{17.7} - 10^{18.7}$ \cmtwo.  Their models
were able to account for polarization strengths as low as 0.002\%,
depending on the total column density of the gas.  The polarization
strengths of the filament stars range from 0.0032\% to 0.1\% (Table
\ref{tab:filament}), however the filament polarizations probably form
over a column that is $< 0.002$ times the length of the column of
interstellar polarizations.  Radiative torques that enhance grain
alignment will become of increasing importance as grains approach the
heliopause and experience a stronger, more asymmetric, and more
energetic radiation field (see Fig. 3 in S12).

As the ISMF strength increases as field is compressed against the
heliopause by the solar motion through the surrounding ISM,
interstellar dust grains will couple more tightly to the ISMF because
of smaller gyroradii.  The interstellar field strength is enhanced by
factors of three or more outside of the heliopause in the models used
by \citet{SlavinFrisch:2012}.  \Voyager\ 1 found that field strengths
in the outer heliosphere reach 6 \microG, which suggests a compression
factor of two when compared to the 3 \microG\ field strength of the
interstellar cloud around the heliosphere
\citep{Burlaga:2014ismf,SlavinFrisch:2008,Schwadronetal:2011sep}.

Polarizations are not additive in the ISM since foreground clouds may
depolarize polarized light
\citep{Serkowski:1975curve,DavisBerge:1968}.  There is no information
on foreground depolarization in the local ISM so the local
significance of this effect is unknown.  However, when compared to
interstellar values, there is unlikely to be foreground depolarization
of polarizations originating in the outer heliosphere.  In principle
the absence of depolarization will further increase the efficiency of
heliosheath polarizations above those in the ISM.

A significant enhancement of polarizations compared to previous
simulations of polarization efficiencies may arise from the physical
structure of the excluded grains, and whether they are compact or
porous aggregates (``fluffy'').  Photoionization models of
  interstellar gas around the heliosphere show the presence of olivine
  silicates and the absence of carbonaceous grains
  \citep{SlavinFrisch:2008}.  The grain trajectory models that predict
  dust plumes around the heliosphere (S12) are based on compact
  olivine silicate grains with densities of 3.3 gr \cc.  
In situ measurements by Stardust found interstellar grains with
very low volume densities, $<0.07$ gr \cc\ \citep{Westphal:2014sci}.
Recently, \citet{Sterken:2015} found an indication of porous
grains from comparisons of the flow directions of interstellar grains
in 16 years of interstellar dust measurements from \Ulysses, such as
was proposed in earlier discussions \citep[cf.][]{Mann:2010araa}.  In
situ measurements of interstellar dust in the inner heliosphere have
found large grains with masses $2 \times 10^{-10}$ gr \citep[radii
  $\sim 2$ \micron\ for compact
  grains,][]{Frisch:1999,SlavinFrisch:2012,Krueger:2014mass}.  These
heliospheric large grains are similar in size to the large
micron-sized aggregate grains that are observed in the cores of dense
interstellar clouds \citep{Pagani:2010dust,Santos:2014}.  Outside of
dense clouds the large grains couple to the interstellar gas over
pathlengths of $\sim 50-100$ pc \citep{GruenLandgraf:2000}, and the
local population of large grains may be enriched by porous dust grains
from fragmented dense clouds.

Porous grains have larger cross-sections than the canonical compact
grains that have been used for modeling the dust-heliosphere
interactions.  \citet{Ma:2013fluffyinism} have modeled the grain
charging of aggregate grains and found that the surface potentials of
porous grains is increased by 30\% or more in the outer heliosheath,
which allows a stronger coupling between the dust and deflected
heliosheath plasma and magnetic field.  Aggregate porous grains have
higher charges than compact grains, and so are deflected differently.
The high end of the cutoff masses for the deflected grains is
increased, so that more high-mass grains are prevented from entering
the heliosphere and the masses of the dust plumes would be enhanced
compared to models based on compact grains.
\citet{HoangLazarian:2014radiativetorques} find that larger grains are
more sensitive to radiative torques, so that the polarizations of
large fluffy grains should be further enhanced over those of compact
grains in the outer heliosheath.

\section{CONCLUDING REMARKS}

A filament of polarized starlight has been discovered within 10 pc of
the heliosphere.  The filament stars were initially identified from
the rotation of polarization position angles with star distance.  This
study links these polarizations to the outer heliosheath regions where
previous simulations indicate that interstellar nanograins are
deflected in plume-like features around the heliosphere.  The filament
polarizations show three primary observational properties: (1) The
best-fitting ISMF vector to the filament polarization position angles
coincides with the upwind velocity vector of interstellar neutrals
flowing into the heliosphere. (2) The axis of this 100\deeg--long
filament is perpendicular to the \BV\ plane, formed by the
interstellar and magnetic and gas-velocity vectors, that shapes the
heliosphere.  (3) A contribution of $ \pm 9.6^\circ$ to the dispersion
of the polarization position angles can be attributed to magnetic
turbulence.

We propose that the filament is formed in the outer heliosheath,
where the ISMF and interstellar plasma are deflected around the
heliosphere.  The polarizations would be found where the flow
streamlines and draped ISMF are approximately parallel to each other,
and roughly perpendicular to the \BV\ plane and sightline.  However,
although the filament traces several physical properties of the outer
heliosheath, the difference in the thickness of the heliosheath
compared to normal interstellar sightlines suggests that grain
alignment mechanisms in the outer heliosheath would have to be more
efficient than in the general ISM for such an origin.

An outer heliosphere origin for the filament can be tested by
evaluating the propagation of porous interstellar dust grains through
MHD models of the dynamically evolving outer heliosheath plasma and
magnetic fields.  Alignment efficiency in the outer heliosheath,
compared to the ISM, would be augmented by increased grain charging,
radiative torques, and stronger magnetic field strengths.  Porous
interstellar grains will substantially increase the mass and number of
deflected interstellar grains above values for compact grains.  The
models will need to keep track of magnetic, collisional, and radiative
torques on the grains in order to evaluate grain alignment and the
asymmetric grain opacity that causes polarization.  Filament
properties that need to be explained include the filament width, the
connection between the filament ISMF vector and neutral He flow
vector, and detailed modeling of grain alignment mechanisms and
polarization efficiencies for porous grains.

The properties of the outer heliosheath are poorly known.  While
\IBEX\ and the two \Voyager\ missions have provided essential data on
the outer heliosphere, data on the vector motion of the outer
heliosheath plasma are not presently available because the \Voyager\ 1
plasma instrument is not operational and \Voyager\ 2 has not yet
reached the heliopause \citep[e.g.][]{RichardsonDecker:2014}.  If an
outer heliosheath origin for the filament polarizations is confirmed
by theoretical modeling, a new diagnostic of the outer heliosheath
will become available.  This diagnostic would have the unique property
of allowing the outer heliosheath to be monitored using ground-based
observations.  This convergence between interstellar dust data and
heliosheath science would be useful for understanding the interactions
between astrospheres around external stars and the ISM,
heliosphere-related foregrounds to the cosmic microwave background,
and the time-variable flux of interstellar grains into the inner
heliosphere and terrestrial atmosphere.

\acknowledgements This research has been partly supported by the NASA
Explorer program through support for the \IBEX\ mission, and by the
European Research Council Advanced Grant HotMol (ERC-2011-AdG 291659).

\begin{thebibliography}{77}
\expandafter\ifx\csname natexlab\endcsname\relax\def\natexlab#1{#1}\fi

\bibitem[{{Andersson} {et~al.}(2014){Andersson}, {Lazarian}, \&
  {Vaillancourt}}]{Andersson:2014araa}
{Andersson}, B.-G., {Lazarian}, A., \& {Vaillancourt}, J.~E. 2014, \araa, {in
  press}

\bibitem[{{Avinash} {et~al.}(2014){Avinash}, {Zank}, {Dasgupta}, \&
  {Bhadoria}}]{AvinashZank:2014instability}
{Avinash}, K., {Zank}, G.~P., {Dasgupta}, B., \& {Bhadoria}, S. 2014, \apj,
  791, 102

\bibitem[{{Bailey} {et~al.}(2010){Bailey}, {Lucas}, \&
  {Hough}}]{planetpol:2010}
{Bailey}, J., {Lucas}, P.~W., \& {Hough}, J.~H. 2010, \mnras, 405, 2570

\bibitem[{{Belheouane} {et~al.}(2012){Belheouane}, {Zaslavsky}, {Meyer-Vernet},
  {Issautier}, {Mann}, \& {Maksimovic}}]{Belheouane:2012stereodust}
{Belheouane}, S., {Zaslavsky}, A., {Meyer-Vernet}, N., {Issautier}, K., {Mann},
  I., \& {Maksimovic}, M. 2012, \solphys, 281, 501

\bibitem[{{Borovikov} \&
  {Pogorelov}(2014)}]{BorovikovPogorelov:2014instability}
{Borovikov}, S.~N. \& {Pogorelov}, N.~V. 2014, \apjl, 783, L16

\bibitem[{{Burlaga} \& {Ness}(2014)}]{Burlaga:2014ismf}
{Burlaga}, L.~F. \& {Ness}, N.~F. 2014, \apj, 784, 146

\bibitem[{{Burlaga} {et~al.}(2014){Burlaga}, {Ness}, Florinski, \&
  Heerikhuisen}]{Burlaga:2014ismfStDev}
{Burlaga}, L.~F., {Ness}, N.~F., Florinski, V., \& Heerikhuisen, J. 2014, \apj,
  792, 134

\bibitem[{{Burlaga} {et~al.}(2013){Burlaga}, {Ness}, \&
  {Stone}}]{Burlaga:2013ismf}
{Burlaga}, L.~F., {Ness}, N.~F., \& {Stone}, E.~C. 2013, Science, 1

\bibitem[{{Bzowski} {et~al.}(2014){Bzowski}, {Kubiak}, {Hlond}, {Sokol},
  {Banaszkiewicz}, \& {Witte}}]{Bzowski:2014ulysses}
{Bzowski}, M., {Kubiak}, M.~A., {Hlond}, M., {Sokol}, J.~M., {Banaszkiewicz},
  M., \& {Witte}, M. 2014, ArXiv e-prints

\bibitem[{{Bzowski} {et~al.}(2012){Bzowski}, {Kubiak}, {M{\"o}bius},
  {Bochsler}, {Leonard}, {Heirtzler}, {Kucharek}, {Sok{\'o}{\l}}, {H{\l}ond},
  {Crew}, {Schwadron}, {Fuselier}, \& {McComas}}]{Bzowski:2012isn}
{Bzowski}, M., {Kubiak}, M.~A., {M{\"o}bius}, E., {Bochsler}, P., {Leonard},
  T., {Heirtzler}, D., {Kucharek}, H., {Sok{\'o}{\l}}, J.~M., {H{\l}ond}, M.,
  {Crew}, G.~B., {Schwadron}, N.~A., {Fuselier}, S.~A., \& {McComas}, D.~J.
  2012, \apjs, 198, 12

\bibitem[{{Chalov} {et~al.}(2010){Chalov}, {Alexashov}, {McComas}, {Izmodenov},
  {Malama}, \& {Schwadron}}]{Chalov:2010ribbon}
{Chalov}, S.~V., {Alexashov}, D.~B., {McComas}, D., {Izmodenov}, V.~V.,
  {Malama}, Y.~G., \& {Schwadron}, N. 2010, \apjl, 716, L99

\bibitem[{{Czechowski} \& {Mann}(2003)}]{CzechowskiMann:2003b}
{Czechowski}, A. \& {Mann}, I. 2003, \aap, 410, 165

\bibitem[{{Davis} \& {Berge}(1968)}]{DavisBerge:1968}
{Davis}, L.~J. \& {Berge}, G.~L. 1968, {Evidence for Galactic Magnetic Fields}
  (the University of Chicago Press), 755

\bibitem[{{Frisch} {et~al.}(2010){Frisch}, {Andersson}, {Berdyugin}, {Funsten},
  {Magalhaes}, {McComas}, {Piirola}, {Schwadron}, {Slavin}, \&
  {Wiktorowicz}}]{Frisch:2010ismf1}
{Frisch}, P.~C., {Andersson}, B., {Berdyugin}, A., {Funsten}, H.~O.,
  {Magalhaes}, M., {McComas}, D.~J., {Piirola}, V., {Schwadron}, N.~A.,
  {Slavin}, J.~D., \& {Wiktorowicz}, S.~J. 2010, \apj, 724, 1473

\bibitem[{{Frisch} {et~al.}(2012){Frisch}, {Andersson}, {Berdyugin}, {Piirola},
  {DeMajistre}, {Funsten}, {Magalhaes}, {Seriacopi}, {McComas}, {Schwadron},
  {Slavin}, \& {Wiktorowicz}}]{Frisch:2012ismf2}
{Frisch}, P.~C., {Andersson}, B.-G., {Berdyugin}, A., {Piirola}, V.,
  {DeMajistre}, R., {Funsten}, H.~O., {Magalhaes}, A.~M., {Seriacopi}, D.~B.,
  {McComas}, D.~J., {Schwadron}, N.~A., {Slavin}, J.~D., \& {Wiktorowicz},
  S.~J. 2012, \apj, 760, 106

\bibitem[{{Frisch} {et~al.}(2015){Frisch}, {Andersson}, {Berdyugin}, {Piirola},
  {Funsten}, {Magalhaes}, {Seriacopi}, {McComas}, {Schwadron}, {Slavin}, \&
  {Wiktorowicz}}]{Frisch:2015ismf3}
{Frisch}, P.~C., {Andersson}, B.-G., {Berdyugin}, A., {Piirola}, V., {Funsten},
  H.~O., {Magalhaes}, A.~M., {Seriacopi}, D.~B., {McComas}, D.~J., {Schwadron},
  N.~A., {Slavin}, J.~D., \& {Wiktorowicz}, S.~J. 2015, \apj, $in~preparation$

\bibitem[{{Frisch} {et~al.}(2014){Frisch}, {Berdyugin}, {Funsten}, {Magalhaes},
  {McComas}, {Piirola}, {Schwadron}, \& {Wiktorowicz}}]{Frisch:2014icns}
{Frisch}, P.~C., {Berdyugin}, A., {Funsten}, H.~O., {Magalhaes}, M., {McComas},
  D.~J., {Piirola}, V., {Schwadron}, N.~A., \& {Wiktorowicz}, S.~J. 2014,
  \emph{in press} in the proceedings of the 13th Annual International
  Astrophysics Conference: Voyager, \IBEX\, and the Interstellar Medium.

\bibitem[{{Frisch} {et~al.}(1999){Frisch}, {Dorschner}, {Geiss}, {Greenberg},
  {Gr\"un}, {Landgraf}, {Hoppe}, {Jones}, {Kr{\"{a}}tschmer}, {Linde},
  {Morfill}, {Reach}, {Slavin}, {Svestka}, {Witt}, \& {Zank}}]{Frisch:1999}
{Frisch}, P.~C., {Dorschner}, J.~M., {Geiss}, J., {Greenberg}, J.~M., {Gr\"un},
  E., {Landgraf}, M., {Hoppe}, P., {Jones}, A.~P., {Kr{\"{a}}tschmer}, W.,
  {Linde}, T.~J., {Morfill}, G.~E., {Reach}, W., {Slavin}, J.~D., {Svestka},
  J., {Witt}, A.~N., \& {Zank}, G.~P. 1999, \apj, 525, 492

\bibitem[{{Frisch} {et~al.}(2011){Frisch}, Redfield, \&
  Slavin}]{Frisch:2011araa}
{Frisch}, P.~C., Redfield, S., \& Slavin, J. 2011, \araa, 49

\bibitem[{{Funsten} {et~al.}(2009){Funsten}, {Allegrini}, {Crew}, {DeMajistre},
  {Frisch}, {Fuselier}, {Gruntman}, {Janzen}, {McComas}, {M{\"o}bius},
  {Randol}, {Reisenfeld}, {Roelof}, \& {Schwadron}}]{Funsten:2009sci}
{Funsten}, H.~O., {Allegrini}, F., {Crew}, G.~B., {DeMajistre}, R., {Frisch},
  P.~C., {Fuselier}, S.~A., {Gruntman}, M., {Janzen}, P., {McComas}, D.~J.,
  {M{\"o}bius}, E., {Randol}, B., {Reisenfeld}, D.~B., {Roelof}, E.~C., \&
  {Schwadron}, N.~A. 2009, Science, 326, 964

\bibitem[{{Funsten} {et~al.}(2014){Funsten}, {Bzowski}, {Cai}, {Dayeh},
  {DeMajistre}, {Frisch}, {Heerikhuijsen}, {Higdon}, {Janzen}, {Larsen},
  {Livadiotis}, {McComas}, {M{\"o}bius}, {Reese}, {Roelof}, {Reisenfeld},
  {Schwadron}, \& {Zirnstein}}]{Funsten:2014}
{Funsten}, H.~O., {Bzowski}, M., {Cai}, D.~M., {Dayeh}, M., {DeMajistre}, R.,
  {Frisch}, P.~C., {Heerikhuijsen}, J., {Higdon}, D.~M., {Janzen}, P.,
  {Larsen}, B., {Livadiotis}, G., {McComas}, D.~J., {M{\"o}bius}, E., {Reese},
  C., {Roelof}, E.~C., {Reisenfeld}, D.~B., {Schwadron}, N.~A., \& {Zirnstein},
  E.~J. 2014, \apj, 215, 13

\bibitem[{{Funsten} {et~al.}(2013){Funsten}, {DeMajistre}, {Frisch},
  {Heerikhuijsen}, {Higdon}, {Janzen}, {Larsen}, {Livadiotis}, {McComas},
  {M{\"o}bius}, {Reese}, {Reisenfeld}, {Schwadron}, \&
  {Zirnstein}}]{Funsten:2013}
{Funsten}, H.~O., {DeMajistre}, R., {Frisch}, P.~C., {Heerikhuijsen}, J.,
  {Higdon}, D.~M., {Janzen}, P., {Larsen}, B., {Livadiotis}, G., {McComas},
  D.~J., {M{\"o}bius}, E., {Reese}, C., {Reisenfeld}, D.~B., {Schwadron},
  N.~A., \& {Zirnstein}, E.~J. 2013, \apj, 776, 30

\bibitem[{{Gr{\"u}n} \& {Landgraf}(2000)}]{GruenLandgraf:2000}
{Gr{\"u}n}, E. \& {Landgraf}, M. 2000, \jgr, 105, 10291

\bibitem[{{Gurnett} {et~al.}(2013){Gurnett}, {Kurth}, {Burlaga}, \&
  {Ness}}]{Gurnett:2013sciismvoy}
{Gurnett}, D.~A., {Kurth}, W.~S., {Burlaga}, L.~F., \& {Ness}, N.~F. 2013,
  Science, 341, 1489

\bibitem[{{Haffner} {et~al.}(2009){Haffner}, {Dettmar}, {Beckman}, {Wood},
  {Slavin}, {Giammanco}, {Madsen}, {Zurita}, \&
  {Reynolds}}]{Haffner:2009review}
{Haffner}, L.~M., {Dettmar}, R.~J., {Beckman}, J.~E., {Wood}, K., {Slavin},
  J.~D., {Giammanco}, C., {Madsen}, G.~J., {Zurita}, A., \& {Reynolds}, R.~J.
  2009, Rev. Modern Physics, 81, 969

\bibitem[{{Haverkorn}(2010)}]{Haverkorn:2010magnetoionic}
{Haverkorn}, M. 2010, in Astronomical Society of the Pacific Conference Series,
  Vol. 438, Astronomical Society of the Pacific Conference Series, ed.
  R.~{Kothes}, T.~L. {Landecker}, \& A.~G. {Willis}, 249

\bibitem[{{Heerikhuisen} \& {Pogorelov}(2011)}]{HeerikhuisenPogorelov:2011}
{Heerikhuisen}, J. \& {Pogorelov}, N.~V. 2011, \apj, 738, 29

\bibitem[{{Heerikhuisen} {et~al.}(2010){Heerikhuisen}, {Pogorelov}, {Zank},
  {Crew}, {Frisch}, {Funsten}, {Janzen}, {McComas}, {Reisenfeld}, \&
  {Schwadron}}]{Heerikhuisen:2010ribbon}
{Heerikhuisen}, J., {Pogorelov}, N.~V., {Zank}, G.~P., {Crew}, G.~B., {Frisch},
  P.~C., {Funsten}, H.~O., {Janzen}, P.~H., {McComas}, D.~J., {Reisenfeld},
  D.~B., \& {Schwadron}, N.~A. 2010, \apjl, 708, L126

\bibitem[{{Heerikhuisen} {et~al.}(2014){Heerikhuisen}, {Zirnstein}, {Funsten},
  {Pogorelov}, \& {Zank}}]{Heerikhuisen:2014}
{Heerikhuisen}, J., {Zirnstein}, E.~J., {Funsten}, H.~O., {Pogorelov}, N.~V.,
  \& {Zank}, G.~P. 2014, \apj, 784, 73

\bibitem[{{Heiles}(2000)}]{Heiles:2000}
{Heiles}, C. 2000, \aj, 119, 923

\bibitem[{{Hoang} \& {Lazarian}(2014)}]{HoangLazarian:2014radiativetorques}
{Hoang}, T. \& {Lazarian}, A. 2014, \mnras, 438, 680

\bibitem[{{Hsieh} {et~al.}(2010){Hsieh}, {Giacalone}, {Czechowski},
  {Hilchenbach}, {Grzedzielski}, \& {Kota}}]{Hsieh:2010enahp}
{Hsieh}, K.~C., {Giacalone}, J., {Czechowski}, A., {Hilchenbach}, M.,
  {Grzedzielski}, S., \& {Kota}, J. 2010, \apjl, 718, L185

\bibitem[{{Katushkina} {et~al.}(2014){Katushkina}, {Izmodenov}, {Wood}, \&
  {McMullin}}]{KatushkinaWood:2014ulysses}
{Katushkina}, O.~A., {Izmodenov}, V.~V., {Wood}, B.~E., \& {McMullin}, D.~R.
  2014, ArXiv e-prints

\bibitem[{{Kimura} \& {Mann}(1998)}]{KimuraMann:1998charge}
{Kimura}, H. \& {Mann}, I. 1998, \apj, 499, 454

\bibitem[{{Kimura} \& {Mann}(1999)}]{KimuraMann:1999}
---. 1999, Earth, Planets, and Space, 51, 1223

\bibitem[{{Kimura} {et~al.}(2003){Kimura}, {Mann}, \&
  {Jessberger}}]{KimuraMann:2003vel}
{Kimura}, H., {Mann}, I., \& {Jessberger}, E.~K. 2003, \apj, 582, 846

\bibitem[{{Kr{\"u}ger} {et~al.}(2014){Kr{\"u}ger}, {Strub}, {Sterken}, \&
  {Gr{\"u}n}}]{Krueger:2014mass}
{Kr{\"u}ger}, H., {Strub}, P., {Sterken}, V.~J., \& {Gr{\"u}n}, E. 2014, \apj,
  submitted

\bibitem[{{Kubiak} {et~al.}(2014){Kubiak}, {Bzowski}, {Sok{\'o}{\l}},
  {Swaczyna}, {Grzedzielski}, {Alexashov}, {Izmodenov}, {Moebius}, {Leonard},
  {Fuselier}, {Wurz}, \& {McComas}}]{KubiakBzowski:2014breeze}
{Kubiak}, M.~A., {Bzowski}, M., {Sok{\'o}{\l}}, J.~M., {Swaczyna}, P.,
  {Grzedzielski}, S., {Alexashov}, D.~B., {Izmodenov}, V.~V., {Moebius}, E.,
  {Leonard}, T., {Fuselier}, S.~A., {Wurz}, P., \& {McComas}, D.~J. 2014,
  \apjs, 213, 29

\bibitem[{{Kurth} \& {Gurnett}(2003)}]{KurthGurnett:2003}
{Kurth}, W.~S. \& {Gurnett}, D.~A. 2003, \jgr, 108, 2

\bibitem[{{Lallement} {et~al.}(2010){Lallement}, {Qu{\'e}merais}, {Koutroumpa},
  {Bertaux}, {Ferron}, {Schmidt}, \& {Lamy}}]{Lallement:2010soho}
{Lallement}, R., {Qu{\'e}merais}, E., {Koutroumpa}, D., {Bertaux}, J.,
  {Ferron}, S., {Schmidt}, W., \& {Lamy}, P. 2010, Twelfth International Solar
  Wind Conference, 1216, 555

\bibitem[{{Landgraf} {et~al.}(2000){Landgraf}, {Baggaley}, {Gr{\" u}n}, {Kr{\"
  u}ger}, \& {Linkert}}]{Landgrafetal:2000}
{Landgraf}, M., {Baggaley}, W.~J., {Gr{\" u}n}, E., {Kr{\" u}ger}, H., \&
  {Linkert}, G. 2000, \jgr, 105, 10343

\bibitem[{{Lazarian}(2007)}]{Lazarian:2007rev}
{Lazarian}, A. 2007, \jqsrt, 106, 225

\bibitem[{{Linde} \& {Gombosi}(2000)}]{LindeGombosi:2000}
{Linde}, T.~J. \& {Gombosi}, T.~I. 2000, \jgr, 105, 10411

\bibitem[{{Ma} {et~al.}(2013){Ma}, {Matthews}, {Land}, \&
  {Hyde}}]{Ma:2013fluffyinism}
{Ma}, Q., {Matthews}, L.~S., {Land}, V., \& {Hyde}, T.~W. 2013, \apj, 763, 77

\bibitem[{{Mann}(2010)}]{Mann:2010araa}
{Mann}, I. 2010, \araa, 48, 173

\bibitem[{{Mann} \& {Czechowski}(2004)}]{MannCzechowski:2004}
{Mann}, I. \& {Czechowski}, A. 2004, in AIP Conf. Proc. 719: Physics of the
  Outer Heliosphere, 53--58

\bibitem[{Mathis {et~al.}(1977)Mathis, Rumpl, \& Nordsieck}]{MRN:1977}
Mathis, J.~S., Rumpl, W., \& Nordsieck, K.~H. 1977, \apj, 217, 425

\bibitem[{{McComas} {et~al.}(2009){McComas}, {Allegrini}, {Bochsler},
  {Bzowski}, {Christian}, {Crew}, {DeMajistre}, {Fahr}, {Fichtner}, {Frisch},
  {Funsten}, {Fuselier}, {Gloeckler}, {Gruntman}, {Heerikhuisen}, {Izmodenov},
  {Janzen}, {Knappenberger}, {Krimigis}, {Kucharek}, {Lee}, {Livadiotis},
  {Livi}, {MacDowall}, {Mitchell}, {M{\"o}bius}, {Moore}, {Pogorelov},
  {Reisenfeld}, {Roelof}, {Saul}, {Schwadron}, {Valek}, {Vanderspek}, {Wurz},
  \& {Zank}}]{McComas:2009sci}
{McComas}, D.~J., {Allegrini}, F., {Bochsler}, P., {Bzowski}, M., {Christian},
  E.~R., {Crew}, G.~B., {DeMajistre}, R., {Fahr}, H., {Fichtner}, H., {Frisch},
  P.~C., {Funsten}, H.~O., {Fuselier}, S.~A., {Gloeckler}, G., {Gruntman}, M.,
  {Heerikhuisen}, J., {Izmodenov}, V., {Janzen}, P., {Knappenberger}, P.,
  {Krimigis}, S., {Kucharek}, H., {Lee}, M., {Livadiotis}, G., {Livi}, S.,
  {MacDowall}, R.~J., {Mitchell}, D., {M{\"o}bius}, E., {Moore}, T.,
  {Pogorelov}, N.~V., {Reisenfeld}, D., {Roelof}, E., {Saul}, L., {Schwadron},
  N.~A., {Valek}, P.~W., {Vanderspek}, R., {Wurz}, P., \& {Zank}, G.~P. 2009,
  Science, 326, 959

\bibitem[{{McComas} {et~al.}(2014{\natexlab{a}}){McComas}, {Allegrini},
  {Bzowski}, {Dayeh}, {DeMajistre}, {Funsten}, {Fuselier}, {Gruntman},
  {Janzen}, {Kubiak}, {Kucharek}, {M{\"o}bius}, {Reisenfeld}, {Schwadron},
  {Sok{\'o}{\l}}, \& {Tokumaru}}]{McComas:2014fiveyear}
{McComas}, D.~J., {Allegrini}, F., {Bzowski}, M., {Dayeh}, M.~A., {DeMajistre},
  R., {Funsten}, H.~O., {Fuselier}, S.~A., {Gruntman}, M., {Janzen}, P.~H.,
  {Kubiak}, M.~A., {Kucharek}, H., {M{\"o}bius}, E., {Reisenfeld}, D.~B.,
  {Schwadron}, N.~A., {Sok{\'o}{\l}}, J.~M., \& {Tokumaru}, M.
  2014{\natexlab{a}}, \apjs, 213, 20

\bibitem[{{McComas} {et~al.}(2014{\natexlab{b}}){McComas}, {Bzowski}, {Frisch},
  {Fuselier}, {Kubiak}, {Kucharek}, {Leonard}, {M{\"o}bius}, {Schwadron},
  {Sokol}, {Swaczyna}, \& {Witte}}]{McComas:2014warm}
{McComas}, D.~J., {Bzowski}, M., {Frisch}, P.~C., {Fuselier}, S.~A., {Kubiak},
  M.~A., {Kucharek}, H., {Leonard}, T., {M{\"o}bius}, E., {Schwadron}, N.~A.,
  {Sokol}, J.~M., {Swaczyna}, P., \& {Witte}, M. 2014{\natexlab{b}}, \apj, in
  press

\bibitem[{{McComas} {et~al.}(2014{\natexlab{c}}){McComas}, {Lewis}, \&
  {Schwadron}}]{McComasLewisSchwadron:2014}
{McComas}, D.~J., {Lewis}, W.~S., \& {Schwadron}, N.~A. 2014{\natexlab{c}},
  Reviews of Geophysics

\bibitem[{{M{\"o}bius} {et~al.}(2012){M{\"o}bius}, {Bochsler}, {Bzowski},
  {Heirtzler}, {Kubiak}, {Kucharek}, {Lee}, {Leonard}, {Schwadron}, {Wu},
  {Fuselier}, {Crew}, {McComas}, {Petersen}, {Saul}, {Valovcin}, {Vanderspek},
  \& {Wurz}}]{Moebius:2012isn}
{M{\"o}bius}, E., {Bochsler}, P., {Bzowski}, M., {Heirtzler}, D., {Kubiak},
  M.~A., {Kucharek}, H., {Lee}, M.~A., {Leonard}, T., {Schwadron}, N.~A., {Wu},
  X., {Fuselier}, S.~A., {Crew}, G., {McComas}, D.~J., {Petersen}, L., {Saul},
  L., {Valovcin}, D., {Vanderspek}, R., \& {Wurz}, P. 2012, \apjs, 198, 11

\bibitem[{{Opher} {et~al.}(2009){Opher}, {Bibi}, {Toth}, {Richardson},
  {Izmodenov}, \& {Gombosi}}]{OpherBibi:2009nature}
{Opher}, M., {Bibi}, F.~A., {Toth}, G., {Richardson}, J.~D., {Izmodenov},
  V.~V., \& {Gombosi}, T.~I. 2009, \nat, 462, 1036

\bibitem[{{Opher} \& {Drake}(2013)}]{OpherDrake:2013}
{Opher}, M. \& {Drake}, J.~F. 2013, \apjl, 778, {L26}

\bibitem[{{Pagani} {et~al.}(2010){Pagani}, {Steinacker}, {Bacmann}, {Stutz}, \&
  {Henning}}]{Pagani:2010dust}
{Pagani}, L., {Steinacker}, J., {Bacmann}, A., {Stutz}, A., \& {Henning}, T.
  2010, Science, 329, 1622

\bibitem[{{Peri} {et~al.}(2012){Peri}, {Benaglia}, {Brookes}, {Stevens}, \&
  {Isequilla}}]{Peri:2012bowshocksurvey}
{Peri}, C.~S., {Benaglia}, P., {Brookes}, D.~P., {Stevens}, I.~R., \&
  {Isequilla}, N.~L. 2012, \aap, 538, A108

\bibitem[{{Piirola} {et~al.}(2014){Piirola}, {Berdyugin}, \&
  {Berdyugina}}]{Piirola:2014spie}
{Piirola}, V., {Berdyugin}, A., \& {Berdyugina}, S. 2014, in Society of
  Photo-Optical Instrumentation Engineers (SPIE) Conference Series, Vol. 9147,
  Society of Photo-Optical Instrumentation Engineers (SPIE) Conference Series,
  8

\bibitem[{{Pogorelov} {et~al.}(2009){Pogorelov}, {Heerikhuisen}, {Mitchell},
  {Cairns}, \& {Zank}}]{Pogorelov:2009ribbon}
{Pogorelov}, N.~V., {Heerikhuisen}, J., {Mitchell}, J.~J., {Cairns}, I.~H., \&
  {Zank}, G.~P. 2009, \apjl, 695, L31

\bibitem[{{Pogorelov} {et~al.}(2011){Pogorelov}, {Heerikhuisen}, {Zank},
  {Borovikov}, {Frisch}, \& {McComas}}]{PogorelovFrisch:2011}
{Pogorelov}, N.~V., {Heerikhuisen}, J., {Zank}, G.~P., {Borovikov}, S.~N.,
  {Frisch}, P.~C., \& {McComas}, D.~J. 2011, \apj, 742, 104

\bibitem[{{Richardson} \& {Decker}(2014)}]{RichardsonDecker:2014}
{Richardson}, J.~D. \& {Decker}, R. 2014, \apj, 792, 126

\bibitem[{{Santos} {et~al.}(2011){Santos}, {Corradi}, \& {Reis}}]{Santos:2010}
{Santos}, F.~P., {Corradi}, W., \& {Reis}, W. 2011, \apj, 728, 104

\bibitem[{{Santos} {et~al.}(2014){Santos}, {Franco}, {Roman-Lopes}, {Reis}, \&
  {Rom{\'a}n-Z{\'u}{\~n}iga}}]{Santos:2014}
{Santos}, F.~P., {Franco}, G.~A.~P., {Roman-Lopes}, A., {Reis}, W., \&
  {Rom{\'a}n-Z{\'u}{\~n}iga}, C.~G. 2014, \apj, 1

\bibitem[{{Schwadron} {et~al.}(2011){Schwadron}, {Allegrini}, {Bzowski},
  {Christian}, {Crew}, {Dayeh}, {DeMajistre}, {Frisch}, {Funsten}, {Fuselier},
  {Goodrich}, {Gruntman}, {Janzen}, {Kucharek}, {Livadiotis}, {McComas},
  {Moebius}, {Prested}, {Reisenfeld}, {Reno}, {Roelof}, {Siegel}, \&
  {Vanderspek}}]{Schwadronetal:2011sep}
{Schwadron}, N.~A., {Allegrini}, F., {Bzowski}, M., {Christian}, E.~R., {Crew},
  G.~B., {Dayeh}, M., {DeMajistre}, R., {Frisch}, P., {Funsten}, H.~O.,
  {Fuselier}, S.~A., {Goodrich}, K., {Gruntman}, M., {Janzen}, P., {Kucharek},
  H., {Livadiotis}, G., {McComas}, D.~J., {Moebius}, E., {Prested}, C.,
  {Reisenfeld}, D., {Reno}, M., {Roelof}, E., {Siegel}, J., \& {Vanderspek}, R.
  2011, \apj, 731, 56

\bibitem[{{Schwadron} {et~al.}(2009){Schwadron}, {Bzowski}, {Crew}, {Gruntman},
  {Fahr}, {Fichtner}, {Frisch}, {Funsten}, {Fuselier}, {Heerikhuisen},
  {Izmodenov}, {Kucharek}, {Lee}, {Livadiotis}, {McComas}, {Moebius}, {Moore},
  {Mukherjee}, {Pogorelov}, {Prested}, {Reisenfeld}, {Roelof}, \&
  {Zank}}]{Schwadron:2009sci}
{Schwadron}, N.~A., {Bzowski}, M., {Crew}, G.~B., {Gruntman}, M., {Fahr}, H.,
  {Fichtner}, H., {Frisch}, P.~C., {Funsten}, H.~O., {Fuselier}, S.,
  {Heerikhuisen}, J., {Izmodenov}, V., {Kucharek}, H., {Lee}, M., {Livadiotis},
  G., {McComas}, D.~J., {Moebius}, E., {Moore}, T., {Mukherjee}, J.,
  {Pogorelov}, N.~V., {Prested}, C., {Reisenfeld}, D., {Roelof}, E., \& {Zank},
  G.~P. 2009, Science, 326, 966

\bibitem[{{Schwadron} \& {McComas}(2013)}]{SchwadronMcComas:2013rr}
{Schwadron}, N.~A. \& {McComas}, D.~J. 2013, \apj, 764, 92

\bibitem[{{Schwadron} {et~al.}(2014){Schwadron}, {Moebius}, {Fuselier},
  {McComas}, {Funsten}, {Janzen}, {Reisenfeld}, {Kucharek}, {Lee}, {Fairchild},
  {Allegrini}, {Dayeh}, {Livadiotis}, {Reno}, {Bzowski}, {Sok\'ol}, {Kubiak},
  {Christian}, {DeMajistre}, {Frisch}, {Galli}, {Wurz}, \&
  {Gruntman}}]{Schwadron:2014sep}
{Schwadron}, N.~A., {Moebius}, E., {Fuselier}, S.~A., {McComas}, D.~J.,
  {Funsten}, H.~O., {Janzen}, P., {Reisenfeld}, D., {Kucharek}, H., {Lee},
  M.~A., {Fairchild}, K., {Allegrini}, F., {Dayeh}, M., {Livadiotis}, G.,
  {Reno}, M., {Bzowski}, M., {Sok\'ol}, J.~M., {Kubiak}, M.~A., {Christian},
  E.~R., {DeMajistre}, R., {Frisch}, P., {Galli}, A., {Wurz}, P., \&
  {Gruntman}, M. 2014, \apjs, 215, 13

\bibitem[{{Serkowski} {et~al.}(1975){Serkowski}, {Mathewson}, \&
  {Ford}}]{Serkowski:1975curve}
{Serkowski}, K., {Mathewson}, D.~S., \& {Ford}, V.~L. 1975, \apj, 196, 261

\bibitem[{{Slavin} \& {Frisch}(2008)}]{SlavinFrisch:2008}
{Slavin}, J.~D. \& {Frisch}, P.~C. 2008, \aap, 491, 53

\bibitem[{{Slavin} {et~al.}(2012){Slavin}, {Frisch}, {M{\"u}ller},
  {Heerikhuisen}, {Pogorelov}, {Reach}, \& {Zank}}]{SlavinFrisch:2012}
{Slavin}, J.~D., {Frisch}, P.~C., {M{\"u}ller}, H.-R., {Heerikhuisen}, J.,
  {Pogorelov}, N.~V., {Reach}, W.~T., \& {Zank}, G. 2012, \apj, 760, 46

\bibitem[{{Spangler} {et~al.}(2011){Spangler}, {Savage}, \&
  {Redfield}}]{SpanglerSavage:2011aip}
{Spangler}, S.~R., {Savage}, A.~H., \& {Redfield}, S. 2011, in American
  Institute of Physics Conference Series, Vol. 1366, American Institute of
  Physics Conference Series, ed. V.~{Florinski}, J.~{Heerikhuisen}, G.~P.
  {Zank}, \& D.~L. {Gallagher}, 97--106

\bibitem[{{Sterken} {et~al.}(2012){Sterken}, {Altobelli}, {Kempf}, {Schwehm},
  {Srama}, \& {Gr{\"u}n}}]{Sterken:2012}
{Sterken}, V.~J., {Altobelli}, N., {Kempf}, S., {Schwehm}, G., {Srama}, R., \&
  {Gr{\"u}n}, E. 2012, \aap, 538, A102

\bibitem[{{Sterken} {et~al.}(2015){Sterken}, {Strub}, {Krueger}, \& {von
  Steiger}}]{Sterken:2015}
{Sterken}, V.~J., {Strub}, P., {Krueger}, H., \& {von Steiger}, R. 2015, in
  preparation, 00

\bibitem[{{Westphal} \& the Stardust~Team(2014)}]{Westphal:2014sci}
{Westphal}, A.~J. \& the Stardust~Team. 2014, Science, 345, 786,791

\bibitem[{{Witte}(2004)}]{Witte:2004}
{Witte}, M. 2004, \aap, 426, 835

\bibitem[{{Wood} {et~al.}(2000){Wood}, {Linsky}, \& {Zank}}]{Wood36Oph:2000}
{Wood}, B.~E., {Linsky}, J.~L., \& {Zank}, G.~P. 2000, \apj, 537, 304

\bibitem[{{Wood} {et~al.}(2014){Wood}, {Mueller}, \&
  {Witte}}]{Wood:2014ulysses}
{Wood}, B.~E., {Mueller}, H.-R., \& {Witte}, M. 2014, \apj, in press

\bibitem[{{Zank} {et~al.}(2013){Zank}, {Heerikhuisen}, {Wood}, {Pogorelov},
  {Zirnstein}, \& {McComas}}]{Zank:2013}
{Zank}, G.~P., {Heerikhuisen}, J., {Wood}, B.~E., {Pogorelov}, N.~V.,
  {Zirnstein}, E., \& {McComas}, D.~J. 2013, \apj, 763, 20

\end{thebibliography}

\begin{deluxetable}{lc rcccc cl}
\tablecaption{Stars tracing magnetic ``filament'' \label{tab:filament} }
\tablewidth{0pt} 
\tabletypesize{\small}
\footnotesize{\tiny}
\tablehead{ 
\colhead{Star} & \colhead{} & \colhead{Coordinates} & \colhead{Distance} & \colhead{Spec.} & \colhead{Polarization\tablenotemark{A}} & \colhead{\PAgal} & \colhead{Ref.\tablenotemark{B}} \\
\colhead{} & \colhead{} & \colhead{$\ell,b$ (deg)} & \colhead{(pc)} & \colhead{} & \colhead{($10^{-5}$ )} & \colhead{(deg)} \\
}
\startdata 
\multicolumn{8}{l}{\emph{Data used in Paper III\tablenotemark{C}}} \nl
HD 131977 & \nodata & 338, 34 & 6 & K4 V & $ 55.4 \pm 24.0 $ & $ 99.2 \pm 12.3$ & LNA \nl  
HD 161797 &  \nodata &   52, 26 &   8  & G5 IV &   $0.93 \pm 0.21$  & $ 92.3 \pm 6.7 $ & PP \nl   
HD 172167   &  $\alpha$ Lyr  &    67,  19   &  8  &  A0 Va & $  1.72 \pm 0.1$ &  $104.7 \pm  1.4$  & PP, W  \nl  
HD 159561   &  $\alpha$ Oph  &  36,  23  &  14  & A5 III V &     $2.34 \pm  0.2 $  & $ 96.4 \pm 2.4$ & PP, W  \nl  
HD 120467 & \nodata & 320, 38  & 14 & K1 III &  $77.0 \pm  23.0 $ &  $120.9 \pm 8.6$  &  San  \nl  
HIP 82283 & \nodata &  4, 18 &  18   & G6 III  &   $113.0  \pm 25.0$ & $ 100.1 \pm 6.3 $ &  San  \nl   
HD 144253 & \nodata & 353, 23 & 19  & K3 V &  $ 13.0 \pm 3.0$  &$  115.2 \pm 7.5$  &  NOT \nl  
HD 119756 & \nodata & 316, 29 & 19  & F2 V & $90.0 \pm 35.0 $  & $122.4 \pm 10.6$ &  Hls  \nl  
HD 130819 & \nodata & 340, 38  & 24 & F3 V & $ 6.0 \pm 3.0 $         & $ 114.9 \pm 14.5$ & NOT \nl  
HD 134987 & \nodata & 339, 27 & 26 & G6 IV-V & $ 14.0 \pm 2.0 $ &$  100.6 \pm 4.5$  &  NOT  \nl  
HD 136894 & \nodata & 340, 24 & 28  & G8 V &  $ 14.0 \pm 4.0 $ &$  106.6 \pm 8.0$  &  NOT \nl  
HD 161096  &  \nodata  &     29, 17  & 25 &  K2 III &   $ 3.2 \pm  0.2 $ &  $ 90.9 \pm 1.9 $  & PP, W  \nl  
HD 161868  &  $\gamma$ Oph  &   28,  15  &    29  & A1 V &     $4.1 \pm 0.3 $ & $ 91.1 \pm 2.1 $ & PP, W  \nl  
\multicolumn{8}{l}{\emph{New data from DiPol-2 at UH88\tablenotemark{D}} } & \nl
HD 153631 & \nodata &   7, 17 &  26 & G0 V & $  9.0 \pm   0.9$ &  $ 59.8 \pm   3.1 $  &  UH\tablenotemark{E}  \nl    
HD 141937 &\nodata &  352, 27 &  33 & G2 V & $  7.9 \pm   1.0$ &  $ 137.0 \pm   4.3 $  &   UH  \nl 
HD 145518 &\nodata &  356, 24 &  33 & G0.5 V & $  5.2 \pm   1.1$ &  $ 5.2 \pm   1.1 $  &   UH   \nl  
\enddata 
\footnotesize{\tiny}
\tablenotetext{A}{The units are degree of polarization.}
\tablenotetext{B}{  References:  LNA--Data collected at LNA Picos dos Dios and listed in
\citet{Frisch:2012ismf2} or \citet{Frisch:2015ismf3}; 
W--Wiktorowicz data from Lick observatory (2014, private communication); 
PP--PlanetPol data from \citet{planetpol:2010}; 
NOT--Paper II;
San--LNA Picos dos Dios data from \citet{Santos:2010}; Hls--Compiled data from different sources in \citet{Heiles:2000}.; UH--Data collected with the UH88 and presented in this paper}
\tablenotetext{C}{\citet{Frisch:2015ismf3}} 
\tablenotetext{D}{These data were obtained with DiPol-2 on the University of Hawaii 2.2m telescope in June and July 2014.}
\tablenotetext{E}{HD 153631 was also observed by NOT in 2010 \citep{Frisch:2012ismf2}, giving 
\Pol$= 5\pm 3 \times 10^{-5}$ degree of polarization and \PAgal$= 61.2^\circ \pm 13.40^\circ $ }
\end{deluxetable}

\end{document}